\documentclass[amsmath,amssymb,aps,pre,twocolumn,superscriptaddress]{revtex4-2}
\usepackage{amsmath,amsfonts}
\usepackage{soul}
\usepackage{dsfont}
\usepackage{graphicx}
\usepackage{bm}
\usepackage{hyperref}
\usepackage{url}
\usepackage[english]{babel}
\usepackage[utf8x]{inputenc}
\usepackage[T1]{fontenc}

\usepackage{color}
\definecolor{darkgreen}{rgb}{0,0.6,0}
\definecolor{darkblue}{rgb}{0,0,0.6}
\definecolor{darkred}{rgb}{0.6,0,0}
\definecolor{darkpurple}{rgb}{0.5,0,0.5}

\newcommand{\mi}{\mathrm{i}}

\def\fig#1{{Fig.~\ref{#1}}}     \def\eq#1{{Eq.~\eqref{#1}}}

\begin{document}

\title{Noise-induced transitions from contractile to extensile active stress in isotropic fluids}

\author{Mathieu Dedenon}
\email{Mathieu.Dedenon@unige.ch}
\affiliation{Department of Biochemistry, University of Geneva, 1211 Geneva, Switzerland}
\affiliation{Department of Theoretical Physics, University of Geneva, 1211 Geneva, Switzerland}

\author{Karsten Kruse}
\email{Karsten.Kruse@unige.ch}
\affiliation{Department of Biochemistry, University of Geneva, 1211 Geneva, Switzerland}
\affiliation{Department of Theoretical Physics, University of Geneva, 1211 Geneva, Switzerland}

\date{\today}
\begin{abstract}
Tissues of living cells are a prime example of active fluids. There is experimental evidence that tissues generate extensile active stress even though their constituting cells are contractile. Fluctuating forces that could result from cell-substrate interactions have been proposed to be able to induce a transition from contractile to extensile active stress. Through analytic calculations and numerical computations, we show that in isotropic active fluids, nonlinearities and a coupling between fluctuating forces and fluid density are necessary for such a transition to occur. Here, both transitions from extensile to contractile and \textit{vice versa} are possible.  
\end{abstract}

\maketitle

\section{Introduction}

Active matter locally extracts energy from its environment to produce mechanical work~\cite{Marchetti2013}. Specifically, activity can result in self-propulsion of particles~\cite{Vicsek1995,Chate2020} or in active mechanical stress~\cite{Kruse2000,Joanny2015}. Biology provides fascinating examples of active matter, for example, in form of the cytoskeleton, which can be found in essentially all living cells. It is a network of filamentous polymers and contains molecular motors that can induce relative displacements between filaments~\cite{Takiguchi1991,Sanchez2012}. In animal cells, it plays a crucial role in cell division and locomotion. On larger scales, tissues of living cells provide another prominent example of
active matter~\cite{Saw2017,Duclos2018,Sano2017}. Tissues can simultaneously exhibit active stress as well as self-propulsion of their constituents~\cite{BlanchMercader2021a}.

On a molecular level, the generation of mechanical stress or self-propulsion is intimately linked to anisotropies. These can extend to larger scales, such that other than 
being isotropic, active matter can exhibit macroscopic nematic, polar, or hexatic orientational order. Specific examples include the actomyosin cortex of suspended cells~\cite{Grill2016,Fritzsche2016} or endothelial cells~\cite{Nelson2005} (isotropic), human pigment and bone cells~\cite{Gruler2000} and muscle precursor cells~\cite{Duclos2014} (nematic), confined muscle precursor cells~\cite{Guillamat2022} and epithelial cells~\cite{Trepat2020} (polar), and epithelial cells~\cite{Eckert2023} (hexatic).

Active stress can be contractile or extensile. We give a precise definition of these notions below. For the time being, it suffices to state that a contractile (extensile) material tends to pull (push) on its local environment. How net contractile or extensile stress is generated in the cytoskeleton or tissues, remains an open question~\cite{Belmonte:2017fs,Lenz2020}. Individual cells typically exhibit contractile behavior, for example, C2C12 cells~\cite{Bruyre2019}. On the contrary, under geometric confinement, tissues of these cells display features of extensile systems~\cite{Duclos2018,Guillamat2022}. A similar phenomenon has been reported for MDCK cells~\cite{Ladoux2021}. Possible solutions to this conundrum involve a competition between adhesion and stress fiber machineries~\cite{Ladoux2021}, or a misalignment between stress and cell shape orientation~\cite{Yeomans2023}.

The presence of fluctuating forces has also been invoked to explain a switch from contractile to extensile behavior~\cite{Vafa2021,Killeen2022,Zhang2023}. In the case of cell monolayers, such forces could result from transient episodes during which the cells move directionally. When both, nonlinearity and noise anisotropy are considered, the active stress is effectively renormalized by fluctuations~\cite{Vafa2021}. By analyzing a linear theory, it has been argued that a fluctuation-induced transition from contractile to extensile active stress is universal~\cite{Killeen2022}. 

In the present work, we first define contractile and extensile active stress in general. Then, we investigate the influence of force 
fluctuations on active stress for compressible isotropic active fluids. Contrary to incompressible active fluids, they allow us to assess the effect of force fluctuations by considering their effect on the contraction instability. We show that, in this case, density-independent force fluctuations do not induce a switch from contractile to extensile active stress. In contrast, density-dependent force fluctuations shift the critical activity necessary for contraction, and can be interpreted as effectively changing the amplitude of the active contractile stress. Finally, we discuss the relation between active stress and active traction forces and how to characterize a noise-induced transition between an extensile and a contractile state.

\section{Contractile and extensile active matter}
\label{sec:contractileExtensile}

In the following, we define the notions of contractile and extensile active fluids. We take a macroscopic point of view and treat the fluid as a continuum in space. Its mechanical state is given by the total stress tensor $\bm{\sigma}_{\rm tot}$. If the fluid is at rest, then at each point $\bm\nabla\cdot\bm{\sigma}_{\rm tot}+\mathbf{f}_{\rm ext}=\bm 0$, where $\mathbf{f}_{\rm ext}$ is an externally applied volume force density.

For an isotropic fluid at rest, the total stress tensor can be expressed by a scalar $\sigma_{\rm tot}$. This quantity can be separated into an active and a passive part $\sigma_{\rm tot}= \sigma_{\rm a}+\sigma_{\rm p}$. Here, $\sigma_{\rm p}=-P=-f+n\mu$, where $P$ is the hydrostatic pressure, $f$ is the free energy density, $n$ the fluid density, and $\mu=\partial f/\partial n$ is the chemical potential. In contrast, $\sigma_{\rm a}$ is generated by local processes within the material through the transformation into mechanical work of chemical or other forms of energy. 

The total stress is defined only up to an additive constant. We fix the constant by requiring $\sigma_{\rm a}=0$ in absence of the internal stress-generating processes. Specifically, for cellular processes driven by the hydrolysis of adenosine-triphosphate (ATP), we take $\sigma_{\rm a}=\zeta\Delta\mu$, where $\Delta\mu=\mu_\mathrm{\rm ATP}-\mu_\mathrm{\rm ADP}-\mu_\mathrm{\rm Pi}$ is the difference in chemical potentials of ATP and its hydrolysis products adenosine-diphosphate (ADP) and inorganic phosphate (P$_i$). At thermodynamic equilibrium, $\Delta\mu=0$ and hence $\sigma_{\rm a}=0$. We define an isotropic active fluid to be contractile if $\sigma_{\rm a}>0$ and extensile in the opposite case.

Consider the hypothetical case where the total stress is active and constant throughout a volume $\Omega$ that contains the fluid. To maintain the fluid at rest at any point on the surface $\partial\Omega$, an external surface force density $\mathbf{t}_{\rm ext}$ must be present such that $\bm{\sigma}_{\rm a}\cdot\bm{\nu}=\mathbf{t}_{\rm ext}$, where $\bm{\nu}$ is the outward normal of the surface $\partial\Omega$. Equivalently for an isotropic fluid, $\bm{\nu}\cdot\mathbf{t}_{\rm ext}=\sigma_{\rm a}$. For a contractile fluid, this expression is positive, such that the external force needs to pull on the surface to maintain mechanical equilibrium. The opposite holds for an extensile active fluid. This is in agreement with the intuitive notion of contractile and extensile materials.   

Note that in mechanical equilibrium, $\bm\sigma_{\rm tot}\cdot\bm{\nu}=\mathbf{t}_{\rm ext}$. However, this relation cannot be used to define the fluid as being contractile or extensile as it is a material property, independent of the specific surface force $\mathbf{t}_{\rm ext}$ applied on $\partial\Omega$. 

We now consider the case of anisotropic active stress. We start with active stress generating processes that exhibit polar orientational order, which is captured by a vector field $\mathbf{p}$. A uniaxial active stress along $\mathbf{p}$ can be written as $\bm{\sigma}_{\rm a}=\sigma_{\rm a}\mathbf{p}\otimes\mathbf{p}$, and the material is called contractile if it is contractile along the direction $\mathbf{p}$, that is,
\begin{align}
\mathbf{p}\cdot\bm{\sigma}_{\rm a}\cdot\mathbf{p}&>0.\label{eq:contractileAnisotropic}
\end{align}
More generally, the active stress from a polar material can be decomposed into a traceless and an isotropic part as $\bm{\sigma}_{\rm a}=\sigma_{\rm a}(\mathbf{p}\otimes\mathbf{p}-(p^2/d)\mathds{1})+\sigma_{\rm a}'\mathds{1}$ in $d$ spatial dimensions, where $\mathds{1}$ is the identity tensor. One recovers the uniaxial active stress for $\sigma_{\rm a}'=\sigma_{\rm a} p^2/d$. In general, one needs to separately analyse the active stress from both contributions. In principle, $\sigma_{\rm a}\sigma_{\rm a}'<0$ is possible, such that the fluid is neither contractile nor extensile. 

In general, several processes can contribute to the generation of anisotropic active stress, like microtubule and actin networks interacting with cell shape. This can translate into different polar fields $\mathbf{p}_i$ having their own dynamics and activities~\cite{Yeomans2023}. One can expand the active stress as different contributions from each individual field, $\bm{\sigma}_{\rm a}=\sum_i\,\sigma_{a,i}(\mathbf{p}_i\otimes\mathbf{p}_i-(p_i^2/d)\mathds{1})+\sigma_{\rm a}'\mathds{1}$. As before, contractility or extensility can be defined at the level of individual polar fields for $\sigma_{a,i}$ and for $\sigma_{\rm a}'$. Alternatively, one can use the spectral decomposition of the total active stress, $\bm{\sigma}_{\rm a}=\sum_{\alpha=1}^{d}\,\sigma_{\alpha}\mathbf{e}_{\alpha}\otimes\mathbf{e}_{\alpha}$, and identify the dominant active direction $\mathbf{e}_d$ through the eigenvalue with largest absolute value $|\sigma_d|\geq|\sigma_{\alpha}|$. The material is then dominantly
contractile when $\mathbf{e}_d\cdot\bm{\sigma}_{\rm a}\cdot\mathbf{e}_d>0$. Let us point out  that, \textit{a priori}, there are no constraints on the signs of the eigenvalues of $\bm{\sigma}_{\rm a}$.

Some active materials exhibit orientational order with higher symmetry, called $p$-atic active materials~\cite{Giomi2022}, and the notion of contractility can be generalized. The $p$-fold symmetry in orientational order defines $p$ equivalent principal directions $\mathbf{e}_p$, with global order parameter $S_p$. The principal directions can be extracted from a $p$-order tensor construct~\cite{Giomi2022} $\mathbf{Q}_p=\sqrt{2^{p-2}}S_p\,[\![\mathbf{e}_p^{\otimes p}]\!]$, where $\otimes p$ is the $p$-tensorial product on the principal directions and $[\![.]\!]$ makes the tensor symmetric and traceless. An active stress is then defined from this tensorial order parameter as $\bm\sigma_{\rm a}=\sigma_{\rm a}[\mathbf{Q}_p]_{{(p-2)}}$ with $p-2$ index contractions. The material is then called contractile if $\mathbf{e}_p\cdot\bm{\sigma}_{\rm a}\cdot\mathbf{e}_p>0$.

In particular for a nematic material where $p=2$, there are two principal directions $\mathbf{e}_2=\pm\mathbf{n}$ defined by the director $\mathbf{n}$. The order parameter is $S$ and the active stress can be expressed as  $\bm\sigma_{\rm a}=\sigma_{\rm a}\mathbf{Q}_2$  with $\mathbf{Q}_2=S(\mathbf{n}\otimes\mathbf{n}-(1/d)\mathds{1})$. An active nematic material is called contractile when $\mathbf{n}\cdot\bm{\sigma}_{\rm a}\cdot\mathbf{n}>0$~\footnote{Note that for $d=2$, a traceless form of active stress gives two opposite eigenvalues of identical magnitude in orthogonal directions $\mathbf{e}_p$ and $\mathbf{e}_{p,\perp}$. In this case, $\mathbf{e}_p$ can be obtained by considering $d=3$ and taking the limit of a very thin layer of material in one direction.}. Here, we have neglected a possible isotropic component of the active stress as is often done in the literature.

In compressible isotropic materials, density gradients can lead to anisotropic stress, $\bm{\sigma}_{\rm a}=\sigma_{\rm a}\bm{\nabla}n\otimes\bm{\nabla}n$~\cite{Kruse2003,Cates2015}. Defining $\mathbf{e}_n=\bm\nabla n/|\bm\nabla n|$, a material is again called contractile if $\mathbf{e}_n\cdot\bm{\sigma}_{\rm a}\cdot\mathbf{e}_n>0$. Hence for $\sigma_{\rm a}<0$, the material resists density gradients as the passive surface tension effect of model H~\cite{Cates2015}. In addition, one can also use $\bm\sigma_{\rm a}=\sigma_{\rm a}\bm\nabla\otimes\bm\nabla n$. In this case, the analysis of a density bump shows that a contraction instability occurs if $\sigma_{\rm a}<0$. At first sight, this might seem to contradict condition~\eq{eq:contractileAnisotropic}. However, it is still true that $\mathbf{e}_p\cdot\bm{\sigma}_{\rm a}\cdot\mathbf{e}_p>0$, where $\mathbf{e}_p$ is the eigenvector associated to the eigenvalue with the largest absolute value of $\bm\nabla\otimes\bm\nabla n$. The negative sign of $\sigma_{\rm a}$ is simply due to the fact that $(\mathbf{e}_p\cdot\bm\nabla)^2 n<0$ for a bump. 

Finally, for polar fluids~\cite{Giomi2008}, an alternative stress tensor is $\bm\sigma_{\rm a}=\sigma_{\rm a}[\bm\nabla\otimes\mathbf{p}+(\bm\nabla\otimes\mathbf{p})^T]$. Again, we diagonalize the tensor $\bm\nabla\otimes\mathbf{p}+(\bm\nabla\otimes\mathbf{p})^T$ and obtain the eigenvector $\mathbf{e}_p$ associated to the eigenvalue with the largest absolute value. The fluid is contractile if $\mathbf{e}_p\cdot\bm{\sigma}_{\rm a}\cdot\mathbf{e}_p>0$. Here, the fluid might be locally contractile or extensile depending on the polar texture.

Whereas the total internal stress of a cell monolayer can be obtained, for example, by inference methods~\cite{Nier.2016} or monolayer stress microscopy~\cite{Tambe2011}, the contractile or extensile nature of nematic active fluids is often deduced from the dynamics of disclination points or topological defects. Indeed, disclination points with a topological charge of +1/2 are polar and typically self-propel in active nematic fluids. In extensile fluids they move in the direction of their ``head'', whereas in contractile fluids they move in the opposite direction. Note that the orientational order parameter relevant in this case is the one appearing in the active stress, such that this proxy has to be taken with care. For example, in tissues, this would be the orientational order of the cytoskeleton. Its orientation might be different from that of other indicators like the distribution of (planar) polarity proteins or the cell shape~\cite{Yeomans2023}, which are often considered as they can be easier to access experimentally than the cytoskeletal orientation.

Special attention has to be paid in a system that interacts with environment through traction forces $\mathbf{f}$, as in the case of self-propelled particles like cells migrating on a substrate. In that case, $\bm\nabla\cdot{\bm\sigma}_\mathrm{\rm tot}+\mathbf{f}_\mathrm{\rm ext}+\mathbf{f}=\bm 0$. In a phenomenological description, the expression of the traction forces can also contain terms that are of the form $\mathbf{f}\sim\bm\nabla\cdot\bm\Sigma$ where $\bm\Sigma$ is a second-rank tensor, for example $\bm\Sigma\sim\mathbf{p}\otimes\mathbf{p}$ for polar materials~\cite{BlanchMercader2021b,Zhao2024}. Within the phenomenological framework it is not possible to distinguish such a term from a contribution to the active stress $\bm\sigma_{\rm a}$. For this, either more microscopic descriptions or measurements~\cite{Sano2024} are required. If not accounted for correctly, this can lead to misclassifications.

It has recently been reported that fluctuating traction forces that do not contain divergence-like terms can on average lead to divergence-like terms. Such terms could effectively generate extensile stress even though the material is contractile~\cite{Vafa2021,Killeen2022}. We study this possibility in the context of an isotropic compressible system.

\section{Scalar active matter in presence of polar fluctuations}

Consider an isotropic active fluid with material density $n$ and let $\mathbf{v}$ denote the corresponding velocity field. In presence of material turnover, the continuity equation can be written as
\begin{align}
\partial_tn+\bm{\nabla}\cdot(n\mathbf{v})&=D\nabla^2n+k_p-k_dn.
\label{eq:continuityEquation}
\end{align}
The source and sink terms with constants $k_p$ and $k_d$ account for turnover of the active material. We restrict our attention to the case of small deviations from the mean density $n_0 = k_p/k_d$, such that higher-order terms in the density are negligible. For the cytoskeleton, $k_p$ and $k_d$ correspond to the polymerization and depolymerization rates, whereas for a tissue they correspond to the rates of proliferation and cell death. The effective diffusion term with constant $D$ accounts, for example, for the effect of (isotropic) noise. 

We focus on the case of low Reynolds number flows that dominate in cytoskeletal and tissue dynamics. Then, momentum conservation is expressed through force balance
\begin{align}
\bm{\nabla}\cdot\bm{\sigma}_\mathrm{tot} = \xi\mathbf{v}+\mathbf{f}.
\label{eq:forceBalanceIsotropic}
\end{align}
Here, the total stress tensor $\bm{\sigma}_\mathrm{tot}$ now also contains a viscous part in addition to the hydrostatic and active parts. Explicitly, we take
\begin{align}
\bm{\sigma}_\mathrm{tot}=2\eta\,\mathbf{S} + \tilde{\eta}\,(\bm{\nabla}\cdot\mathbf{v})\mathds{1}+(\alpha-\beta n^2)n\,\mathds{1}.\label{eq:totalStressIsotropic}
\end{align}
In this expression, $\mathbf{S}=[\![\bm{\nabla}\otimes\mathbf{v}]\!]$ is the symmetric trace-free velocity gradient tensor and $\eta$ and $\tilde{\eta}$ the shear and bulk kinematic viscosities.

The final term in \eq{eq:totalStressIsotropic} comprises active and hydrostatic components. Within a purely phenomenological approach, these two contributions to the stress cannot be disentangled. Often, the lower order term proportional to $\alpha$ is considered to be active, whereas the term proportional to $\beta$ to be hydrostatic. Other combinations of polynomial terms can be chosen and also sigmoidal dependencies of the stress on density have been considered~\cite{Julicher2018}. Here, we choose $\sigma_a=\alpha n$. In light of the discussion in Sect.~\ref{sec:contractileExtensile}, a transition from contractile to extensile or \textit{vice versa} corresponds to a change in the sign of $\alpha$. We take $\beta>0$ to assure stability of the solutions. 

In \eq{eq:forceBalanceIsotropic}, the forces resulting from the material stress are balanced by friction and (fluctuating) traction forces, and we neglect other possible external forces. We assume friction forces to be linear in the velocity $\mathbf{v}$ with friction coefficient $\xi$. This choice is appropriate in the case of an immobile background. In our motivating example of a cell monolayer, this could be a substrate the cells are deposited on. 

The fluctuating traction force density $\mathbf{f}$ is connected to cell migration, which often takes the form of a persistent random walk~\cite{Selmeczi2008}. We assume those forces to be persistent and advected with the flow velocity $\mathbf{v}$~\cite{Vafa2021}
\begin{equation}
\tau_p\partial_t\mathbf{f}+\tau_p(\mathbf{v}\cdot\bm{\nabla})\mathbf{f}=-\mathbf{f}+\bm{\theta}\label{eq:forcePersistent}.
\end{equation}
Here, $\theta$ is a gaussian white noise with $\langle\theta_i(\mathbf{r},t)\rangle=0$ and $\langle\theta_i(\mathbf{r},t)\theta_j(\mathbf{r}',t')\rangle=\Delta\delta_{ij}\delta(\mathbf{r}-\mathbf{r}')\delta(t-t')$ for $i,j=1,\ldots,d$.  The characteristic time $\tau_p$ is related to the persistence time of cell migration. We will consider the cases $\tau_p=0$ and $\tau_p>0$ in turn, where $\tau_p=0$ corresponds to uncorrelated gaussian noise $\mathbf{f}=\bm{\theta}$.

In the following sections, we discuss first the deterministic dynamics $\Delta=0$. We then study the effect of fluctuations on the stress, compute velocity-density gradient correlations, and finally consider persistent density-dependent force fluctuations.

\subsection{Deterministic dynamics}

In absence of fluctuating forces and for an infinite system size, the steady state solution to Eqs.~\eqref{eq:continuityEquation}-\eqref{eq:totalStressIsotropic} is given by $n_0= k_p/k_d$ and $\mathbf{v}_0=0$. Initially small density and velocity fluctuations, $\delta n$ and $\mathbf{v}$, around this state evolve according to the linearized dynamic equations
\begin{align}
\partial_t\delta n-D\nabla^2\delta n+k_d\delta n&=-n_0\bm{\nabla}\cdot\mathbf{v}\label{eq:dtrholin}\\
\eta\nabla^2\mathbf{v}+\left(\tilde{\eta}
+\frac{d-2}{d}\eta\right)\bm{\nabla}(\bm{\nabla}\cdot\mathbf{v})-\xi\mathbf{v}&=-\bar{\alpha}\bm{\nabla}\delta n,\end{align}
where $\bar{\alpha}=\alpha-3\beta n_0^2$. 

Decomposing the density and velocity perturbations into Fourier modes $\mathrm{e}^{\mi\mathbf{q}\cdot\mathbf{x}}$, their evolution with time $t$ is proportional to $\exp\left[s(\mathbf{q})t\right]$, where 
\begin{equation}
s(\mathbf{q})=-k_d-Dq^2+\frac{\bar{\alpha}n_0q^2}{\xi+\bar{\eta} q^2}
\label{eq:growthExponent}
\end{equation}
with $\bar{\eta}=\tilde{\eta}+2(d-1)\eta/d$. For sufficiently large values of $\bar{\alpha}$, a mode $\mathbf{q}$ will grow, indicating an instability of the homogeneous state. The critical value $\bar{\alpha}_c$ such that the homogeneous state is unstable for $\bar{\alpha}>\bar{\alpha}_c$ is
\begin{equation}
\bar{\alpha}_c=\frac{D\xi}{n_0}\left[1+\sqrt{\frac{\bar{\eta} k_d}{ D\xi}}\,\right]^2
\end{equation}
At this critical value, the unstable modes are those with $q_c^2=\sqrt{\xi k_d/(\bar{\eta}D)}$. From the critical value it is obvious that the homogeneous state is stabilized by friction forces, diffusion, and turnover. 

We can get further insight into this instability by considering the case  $\eta=\tilde{\eta}=0$, often referred to as the dry limit. Then, $\xi\mathbf{v}\simeq\bar{\alpha}\bm{\nabla}n$. Inserting this expression for the velocity into \eq{eq:dtrholin} we obtain an effective diffusion constant $D_{\rm eff}=D-\bar{\alpha}n_0/\xi$. At sufficiently large contractility $\alpha$, we have $D_{\rm eff}<0$, which favours density accumulation and thus destabilizes the homogeneous state. 

\begin{figure}
\includegraphics[width=0.49\textwidth]{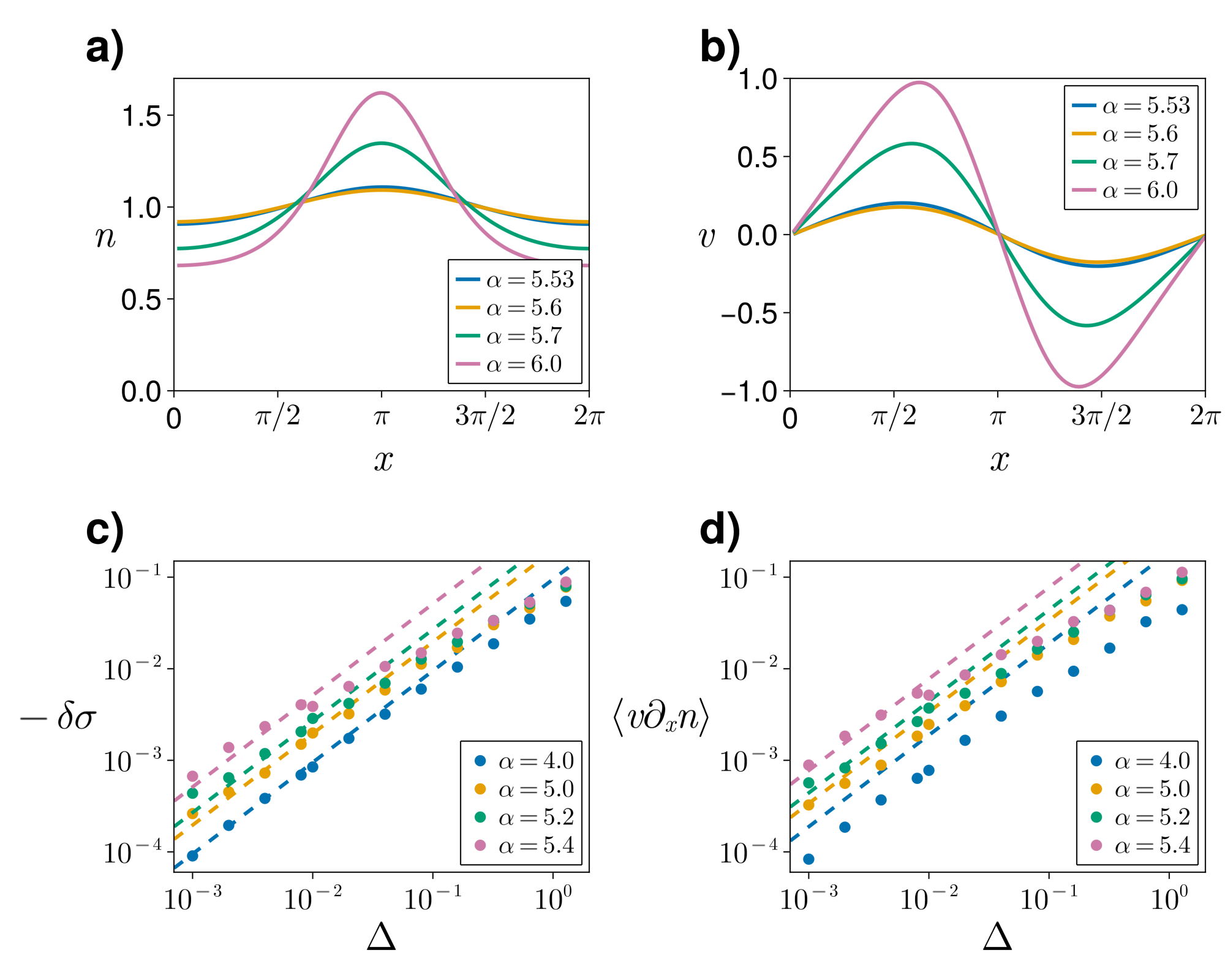}
\caption{Numerical analysis of an isotropic contractile fluid in one spatial dimension with periodic boundary conditions. a) Stationary solutions for a supercritical activity $\alpha>\alpha_c$ in the absence of fluctuating forces.  b) Steady-state velocity profiles corresponding to the states in (a). c) Dependence of the steady-state stress deviation $\delta\sigma=\sigma-\sigma_0$ on the noise amplitude for the homogeneous state (subcritical regime $\alpha<\alpha_c$). Dots: simulations, lines: analytic result Eq.~\eqref{eq:deltaStress} assuming $\langle\delta n^3\rangle=0$. d) As in (c), but for the velocity-density gradient correlation. Dots: simulations, lines: analytic result Eq.~\eqref{eq:vGradnCorr}. Remaining parameters: $\beta=0.5$, $\xi=1$, $D=1$, $\eta=\tilde\eta=1$, $k_p=1$, $k_d=1$, and $L=2\pi$ implying $\alpha_c=5.5$.\label{fig:isotropicOneD}}
\end{figure}

Numerical solutions of the dynamic equations~\eqref{eq:continuityEquation}-\eqref{eq:forcePersistent} in one spatial dimension and for periodic boundary conditions show that beyond the instability, the system develops a region of high density - it contracts, \fig{fig:isotropicOneD}a. The velocity distribution indeed points towards the density maximum, \fig{fig:isotropicOneD}b. The numerical solutions also show that the bifurcation from the homogeneous state is supercritical.

\subsection{Modification of the stress due to fluctuating forces}

In the following, we compute the expectation value of the stress for the homogeneous state in the presence of fluctuating forces for a one-dimensional system under periodic boundary conditions.

We start with the case of uncorrelated fluctuations, $\tau_p=0$. We consider the homogeneous state and assume that the fluctuations $\delta n$ and $\delta\mathbf{v}$ of the density and velocity with respect to this state are small. If the stress were linear in $n$ and $\mathbf{v}$, then there would be no net change in the average stress, $\langle\sigma-\sigma_0\rangle\equiv\langle\delta\sigma\rangle=0$, because $\langle\delta n\rangle=0$ and $\langle\delta \mathbf{v}\rangle=0$ for $f=\theta$. Here, $\sigma_0=(\alpha-\beta n_0^2)n_0$ is the stress in the homogeneous state. 

From expression \eq{eq:totalStressIsotropic} of the stress, we get
\begin{align}
\langle\delta\sigma\rangle &= -3\beta n_0\langle\delta n^2\rangle-\beta\langle\delta n^3\rangle.
\label{eq:deltaStress}
\end{align}
The viscous stress does not contribute, because it is linear in velocity. The first term is negative and proportional to the amplitude $\Delta$ of the fluctuating force. Explicitly in one dimension and  in the limit $k_d\rightarrow 0$, see App.~\ref{app:correlation},
\begin{equation}
\langle\delta n^2\rangle=\frac{\Delta n_0}{4\bar{\alpha}\sqrt{\eta\xi}}\left[-1+\frac{1}{\sqrt{1-\bar{\alpha}/\bar{\alpha}_c)}}\right],
\end{equation}
which is well defined and positive as long as the homogeneous state is linearly stable, that is, $\bar{\alpha}<\bar{\alpha}_c$. The second term vanishes by symmetry because $\langle\delta n\rangle=0$. As a consequence, the stress is expected to decrease linearly with the noise amplitude $\Delta$. Our numerical solutions of Eqs.~\eqref{eq:continuityEquation}-\eqref{eq:forcePersistent} for $\tau_p=0$ and $\Delta>0$ confirm this result for small noise amplitudes $\Delta$, \fig{fig:isotropicOneD}c. It shows that uncorrelated fluctuations do not affect the contraction instability for an isotropic active fluid.

\subsection{Velocity-density gradient correlations}

The results of the previous section show that nonlinearities are in general needed in isotropic systems to modify the stress through fluctuating forces. This is different from Ref.~\cite{Killeen2022}, where a linear theory was used to argue that for anisotropic matter, force fluctuations universally induce a transition from contractile to extensile states. The statement was based on a correlation between the velocity and the nematic tensor $\mathbf{Q}_2$, namely, $\langle\mathbf{v}\cdot(\bm{\nabla}\cdot\mathbf{Q}_2)\rangle$, which was used as a proxy for determining the sign of the active stress. We now compute an analogous correlation for an isotropic active material.

For a compressible isotropic fluid the role of $n$ is somewhat analogous to that of $\mathbf{Q}_2$ for an incompressible nematic fluid, as in the two cases, gradients in $n$ and in $\mathbf{Q}_2$, respectively, give rise to flows. In addition,  fluctuations $\delta n$ and $\delta\mathbf{Q}_2$ around the steady state values relax as $-(q^2+q_0^2)\delta n$ or $-(q^2+q_0^2)\delta \mathbf{Q}_2$ in absence of activity, where $q_0=\sqrt{k_d/D}$ for the isotropic case. For the anisotropic case, $q_0=1/\sqrt{D\tau_Q}$ using the notation given in Ref.~\cite{Vafa2021}. We thus compute the correlation $\langle\mathbf{v}\cdot\bm{\nabla}n\rangle$. 

Following the argument of Ref.~\cite{Killeen2022} for an isotropic active material where activity and friction dominate, we find that $\alpha\langle\mathbf{v}\cdot\bm{\nabla}n\rangle\simeq\xi\langle v^2\rangle\ge0$, such that the sign of the correlation should be the same as the sign of $\alpha$ and thus might be a good proxy for the sign of the active stress. 

Computing the correlation for the fluctuations $\delta n$ and $\delta\mathbf{v}$ around the homogeneous state, we find for uncorrelated gaussian force fluctuations, $\tau_p=0$, and in $d$ spatial dimensions
\begin{equation}
\langle\mathbf{v}\cdot\bm{\nabla}n\rangle=\frac{\pi\,\Delta n_0}{(2\pi)^{d+1}}\int\mathrm{d}^d\mathbf{q}\,\frac{q^2(k_d+Dq^2)}{-s(\mathbf{q})(\xi+\bar{\eta}q^2)^2},
\label{eq:vGradnCorr}
\end{equation}
where $s(\mathbf{q})$ is the growth rate given in \eq{eq:growthExponent}. The derivation is detailed in Appendix~\ref{app:correlation}. This result shows that the sign of the correlation only depends on the sign of the growth exponent $s(\mathbf{q})$ calculated in absence of fluctuations. Since it is strictly negative for a stable homogeneous state, we find a positive correlation $\langle\mathbf{v}\cdot\bm{\nabla}n\rangle$ for any $\alpha<\alpha_c$. In one dimension, one finds $\langle v\partial_x n\rangle = \Delta D/(4\eta\bar{\alpha})\sqrt{\xi/\eta}\,[1-\sqrt{1-\bar{\alpha}/\bar{\alpha}_c}]$ in the limit $k_d\rightarrow 0$, positive-definite for $\bar{\alpha}<\bar{\alpha}_c$. A similar calculation can be performed with persistent noise, see \eq{eq:forcePersistent}, and the result is qualitatively unchanged, App.~\ref{app:correlation}. Our analytic calculations agree well with our numerics, \fig{fig:isotropicOneD}d.

For $\bar{\alpha}<\bar{\alpha}_c$, the velocity-density gradient correlation is positive such that the velocity $\mathbf{v}$ points on average in the same direction as density gradients. According to the arguments given above, one would thus infer a positive value of $\alpha$. However, for $\bar{\alpha}<\bar{\alpha}_c$, the value of $\alpha$ can be both, positive and negative. We conclude that the correlation $\langle\mathbf{v}\cdot\bm{\nabla}n\rangle$ does not indicate whether a compressible isotropic active fluid is contractile or extensile. 

\subsection{Persistent density-dependent force fluctuations}

Up to now, we considered the force fluctuations to be autonomous. In this section, we study the effects of persistent and density-dependent fluctuating forces. Within our approach this amounts to introducing a density-dependent term into~\eq{eq:forcePersistent}. In addition to coupling the force directly to the density, one can also consider a coupling to density gradients, which would be anisotropic. The source of anisotropy of a fluctuating force could also be extrinsic. An example of the latter case is given by a tissue moving on a patterned substrate, for example, by a heterogeneous topography~\cite{Zhao2024,Endresen2021}. In contrast, intrinsic anisotropic force fluctuations could result from cell shape anisotropy~\cite{Zhang2023}.

Explicitly, we modify~\eq{eq:forcePersistent} as follows
\begin{align}
\label{eq:forcePersistentAniso}
\tau_p\left[\partial_t\mathbf{f}+(\mathbf{v}\cdot\bm{\nabla})\mathbf{f}\right]
=-\mathbf{f}+\epsilon_0(n-n_0)\mathbf{f} + \epsilon_1\mathbf{f}\cdot\bm\nabla\bm\nabla n+\bm{\theta},
\end{align}
where $\bm\theta$ is gaussian white noise as before. Let us point out that the amplitude of the gaussian noise is independent of the dynamic fields, which is in contrast to Ref.~\cite{Vafa2021}, where the amplitude is proportional to the expectation value of $\mathbf{Q}_2$. We have neglected a possible linear coupling of $\mathbf{f}$ to density gradients, because it would affect the stability of the homogeneous state even in the absence of fluctuations. The form chosen for~\eq{eq:forcePersistentAniso} assures that the force fluctuations are gaussian for $n=n_0$ and $\tau_p=0$.

We can obtain an expression for $\langle\mathbf{f}\rangle$ following the same procedure as in Ref.~\cite{Vafa2021}, see App.~\ref{app:anisotropiccorrelation}. We obtain
\begin{equation}
\label{eq:meanFdensityDependent}
\langle\mathbf{f}\rangle=\frac{\Delta\, I}{\xi\,l^d}\bm{\nabla}(\epsilon_0n+\epsilon_1\nabla^2n).
\end{equation}
where $l=\sqrt{\bar\eta/\xi}$ is the hydrodynamic length and $I>0$ is a dimensionless number. In the derivation of~\eq{eq:meanFdensityDependent}, we neglected cross-correlations between density and force. Let us also note that the (nonlinear) advection term in~\eq{eq:forcePersistentAniso} is essential for obtaining this result, because otherwise $\langle \theta_i\rangle=0$ implies $\langle f_i\rangle = 0$ even in presence of force-density couplings, see App.~\ref{app:anisotropiccorrelation}.

According to~\eq{eq:meanFdensityDependent}, the average force resulting from fluctuations through~\eq{eq:forcePersistentAniso} is given by the gradient of a density-dependent quantity. In the force-balance equation~\eqref{eq:forceBalanceIsotropic}, this can be interpreted as a modification of the stress. In this case, we see that $\alpha$ is replaced by $\alpha_{\rm eff}=\alpha-\epsilon_0\Delta I/(\xi l^d)$. Moreover, the term proportional to $\epsilon_1$ gives a fluctuation-induced stress of the form $\bm\sigma_1=\kappa\bm\nabla\otimes\bm\nabla n$, as discussed in Sect.~\ref{sec:contractileExtensile}, with $\kappa=-\epsilon_1\Delta I/(\xi l^d)$.

We see that $\epsilon_0$ can change the sign of the active stress $\alpha$, and induce a transition from contractile to extensile or vice-versa. To study the effect of $\epsilon_1$, we perform a linear stability analysis of the deterministic system including the $\bm\sigma_1$ in the total stress, one finds for the growth exponent of the mode $\mathbf{q}$
\begin{equation}
s(\mathbf{q})=-k_d-Dq^2+\frac{(\bar{\alpha}-\kappa q^2)n_0q^2}{\xi+\bar{\eta} q^2}.
\label{eq:growthExponentAniso}
\end{equation}
The critical isotropic activity $\bar{\alpha}_c$ beyond which the homogeneous state becomes unstable then is
\begin{equation}
\bar{\alpha}_c=\frac{D\xi}{n_0}\left[1+2\sqrt{\frac{\bar\eta k_d}{D\xi}\left(1+\frac{\kappa n_0}{\bar\eta D}\right)}+\frac{\bar{\eta} k_d}{ D\xi}\right]
\end{equation}
with the critical wave number $q_c$ satisfying $q_c^2=\sqrt{\xi k_d/(\bar\eta D+\kappa n_0)}$.
This implies an increase of $\bar{\alpha}_c$ for $\kappa>0$ and a decrease in the opposite case. This suggests that density-dependent anisotropic force fluctuations can induce transitions between extensile and contractile fluids with fixed isotropic activity $\alpha$. Note that a stabilizing term $\bm\sigma=-\gamma\bm\nabla\otimes\bm\nabla\nabla^2n$ with $\gamma>0$ should be added to suppress large $q$ instabilities.

\begin{figure}
\includegraphics[width=0.48\textwidth]{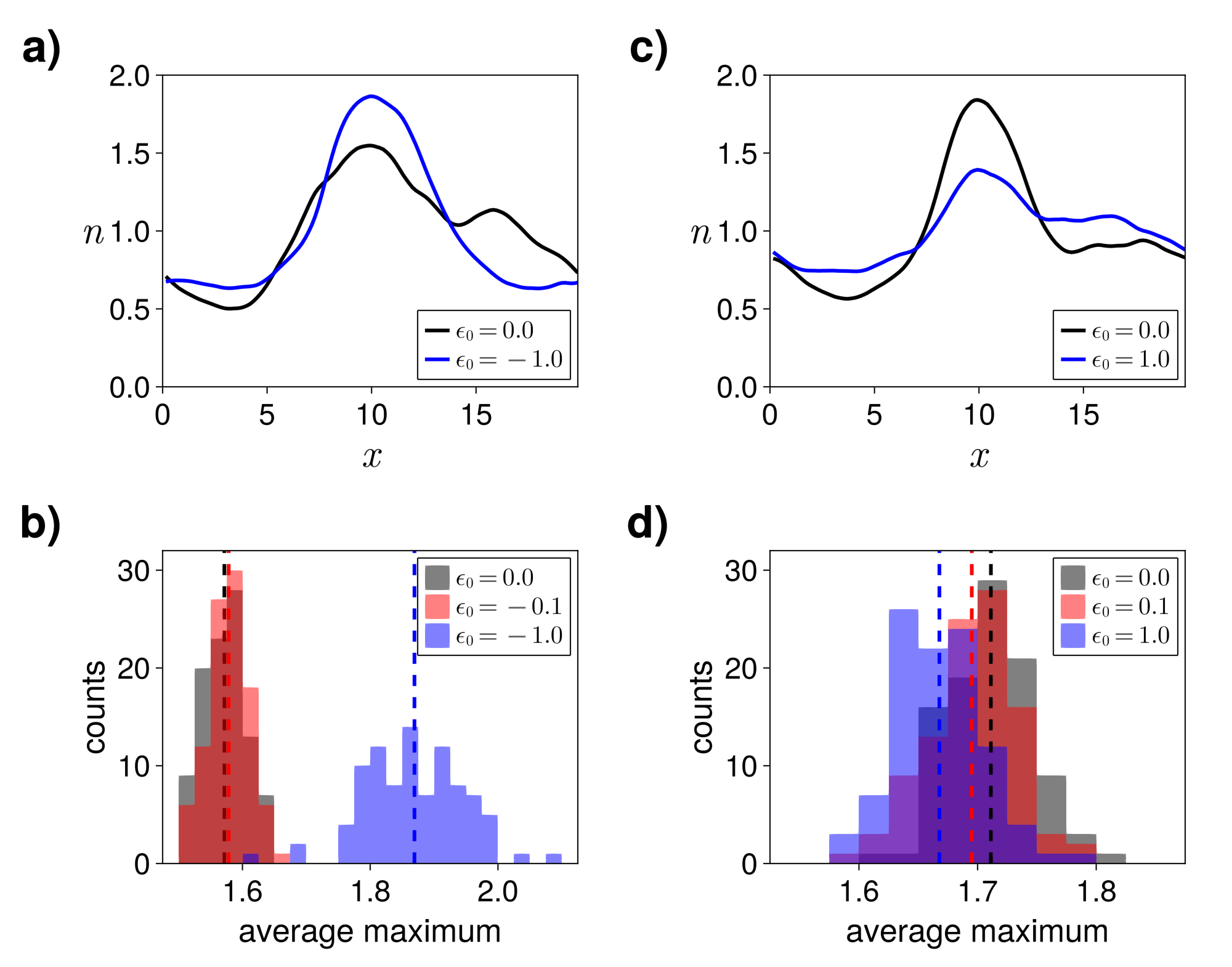}
\caption{Numerical analysis of an isotropic contractile fluid in one spatial dimension with periodic boundary conditions in presence of density-dependent force fluctuations, \eq{eq:forcePersistentAniso}. a) Snapshots of density profiles for a subcritical activity $\alpha=2.1<\alpha_c$ in the presence of fluctuating forces and $\epsilon_0=0$ (black) and $\epsilon_0=1$ (blue). b) Distribution of time-averaged density maxima from 100 numerical solutions as in (a). Averages were taken over a period time $T=100\tau_p$. Dashed lines: distribution means. c) As (a) for a supercritical activity $\alpha=2.3$. d) As in (c) from 100 numerical solutions as in (b). 
Parameters: $\beta=1/3$, $\xi=0.01$, $D=1$, $\eta=\tilde\eta=1$, $k_p=1$, $k_d=1$, $\tau_p=1$, $\Delta=0.1$, and $L=2\pi*\sqrt{10}$. These values imply $\alpha_c\simeq2.21$.  \label{fig:anisotropic}}
\end{figure}

Since the derivation of \eq{eq:meanFdensityDependent} involved some approximations, we numerically checked whether the effective change of $\alpha$ computed above manifests itself in our numerical solutions of the dynamic Eqs.~\eqref{eq:continuityEquation}-\eqref{eq:totalStressIsotropic} and \eqref{eq:forcePersistentAniso}. To have a sizeable modification of $\alpha$ due to fluctuations, we chose $\xi=0.01$ along with $\Delta=0.1$, $\tilde{\eta}=1$. Other parameter values where chosen such that $\alpha_c\simeq2.21$, see caption of Fig.~\ref{fig:anisotropic}. Note that in higher dimensions $d\ge2$ the effects of anisotropic force fluctuations, \eq{eq:meanFdensityDependent}, would be larger for $\xi>1$ rather than for $\xi<1$.

Solving the dynamic equations for a subcritical value of $\alpha=2.1$ and using a negative value of $\epsilon_0$, which implies $\alpha_\mathrm{eff}>\alpha$, we find a distribution that differs from a typical distribution for $\epsilon_0=0$, Fig.~\ref{fig:anisotropic}a. In particular, for $\epsilon_0=-1$, it usually exhibits one well-defined maximum, whereas this is not the case for $\epsilon_0=0$. This result indicates that density-dependent force fluctuations can indeed effectively increase the active stress and induce a contractile instability. Distributions obtained from 100 independent numerical solutions clearly show an increase of the average maximum for $\epsilon_0=-1$ as compared to $\epsilon_0=0$ (two-sided p-value for Welch's unequal variances t-test $p<10^{-60}$), Fig.~\ref{fig:anisotropic}b. No significant change was obtained for $\epsilon_0=-0.1$ compared to $\epsilon_0=0$ ($p=0.13$), Fig.~\ref{fig:anisotropic}b. This is consistent with the corresponding values of $\alpha_\mathrm{eff}$, which we estimate to be $\alpha_\mathrm{eff}\simeq2.11<\alpha_c$ for $\epsilon_0=-0.1$ and $\alpha_\mathrm{eff}\simeq2.22>\alpha_c$ for $\epsilon_0=-1$ when using the value of $I=1/8$ computed in App.~\ref{app:anisotropiccorrelation}.

An intriguing property of the value of $\alpha_\mathrm{eff}$ is its dependence on the sign of $\epsilon_0$,~ \eq{eq:meanFdensityDependent}. For $\epsilon_0>0$, the effective value $\alpha_\mathrm{eff}$ should decrease. We thus also considered the case of $\alpha=2.3>\alpha_c$ and $\epsilon_0>0$. In that case, the differences between instantaneous density profiles for $\epsilon_0=0$, $\epsilon_0=0.1$, and $\epsilon_0=1$ are less pronounced than in the case of $\alpha=2.1$ discussed before, Fig.~\ref{fig:anisotropic}c. However, the distribution of average maxima is clearly shifted to lower values, Fig.~\ref{fig:anisotropic}d, compatible with an effective reduction of $\alpha$. Welch's unequal variances t-test gives a two-sided p-value of $p=0.0016$ for the means of the distributions for $\epsilon_0=0$ and $\epsilon_0=0.1$ to be equal and $p<10^{-14}$ for $\epsilon_0=0$ and $\epsilon_0=1$.

\section{Discussion}

In this work,  using a scalar active system, we show that nonlinearities are essential for a noise-induced transition between extensile and contractile active stress. Specifically, we show that this transition requires (nonlinear) advection and density-dependent force fluctuations. Using a 1d system, we go beyond analytical results by analyzing numerically this behavior. Also, we show that a correlation function similar to the one used in Ref.~\cite{Killeen2022} is not a good quantity to characterise the noise-induced transition. 

The sign of the coefficient $\epsilon_0$ coupling the fluctuating force to the density is not fixed \textit{a priori}, and both a reduction and an augmentation of the active stress can be achieved. However, for cellular tissues, we expect regions of high density, $\delta n\equiv n-n_0>0$, to inhibit polarization and reduce persistence. In \eq{eq:forcePersistentAniso}, an effective persistence time $\tau_{\rm p,eff}=\tau_p/(1-\epsilon_0\,\delta n)$ should decrease with $\delta n>0$. Hence, we expect $\epsilon_0<0$ and thus $\alpha_\mathrm{eff}>\alpha$, interpreted as a noise-induced contraction. Other nonlinear effects might be at play and lead to a different sign of $\epsilon_0$~\cite{Mandal2020}.

In contrast to previous analytical work~\cite{Vafa2021,Killeen2022}, we use the contraction instability as a readout of a noise-induced transition. An important point to emphasize is that active polar fluctuations result from particle-substrate interactions (tractions), whereas an active stress involves particle-particle interactions. Therefore strictly speaking, an effective active stress cannot emerge from polar fluctuations, but the nonlinear couplings between the different fields introduce a phenomenological term that can be interpreted as an effective active stress. An extreme example for an effective stress induced by fluctuations is given by a dry nematic system of self-propelled rods on a substrate, where no particle-particle active force dipoles exist but signatures of extensile nematics are found at high density~\cite{Shi2013,Grossmann2020}. 

The existence of a noise-induced transition between contractile and extensile behavior is in line with previous works for nematic active systems~\cite{Vafa2021,Killeen2022,Zhang2023}. In experiments, however, it remains difficult to distinguish between effects that are due to (active) stress within tissues and those due to traction forces of cells interacting with a substrate. In both cases, fluctuations can qualitatively change the tissue properties. Further experimental efforts are needed to definitely identify the origin of transitions between extensile and contractile activities in confluent tissues.

\begin{acknowledgments}
We thank B. Chakrabarti, C. Blanch-Mercader, S. Henkes, M. Kardar, and T. Liverpool for discussions and C. Blanch-Mercader and Chiu Fan Lee for comments on the manuscript. KK would like to thank the Isaac Newton Institute for Mathematical Sciences, Cambridge, for support and hospitality during the program 'New statistical physics in living matter', where work on this paper was undertaken. This work was supported by EPSRC grant no EP/R014604/1.
\end{acknowledgments}

\appendix

\section{Fourier transforms}
\label{app:Fourier}
In $d$ spatial and $1$ temporal dimensions, one defines the Fourier transform $\mathcal{F}[f]=\tilde{f}$ of a function $f$ and its inverse as
\begin{align}
f(t,\mathbf{x})&=(2\pi)^{-(d+1)/2}\int\mathrm{d}\omega\mathrm{d}^{d}\mathbf{q}\,\tilde{f}(\omega,\mathbf{q})\mathrm{e}^{\mi\omega t+\mi\mathbf{q}\cdot\mathbf{x}} \\
\tilde{f}(\omega,\mathbf{q})&=(2\pi)^{-(d+1)/2}\int\mathrm{d}t\mathrm{d}^{d}\mathbf{x}\,f(t,\mathbf{x})\mathrm{e}^{-\mi\omega t-\mi\mathbf{q}\cdot\mathbf{x}}
\end{align}
with
\begin{align}
(2\pi)^{d+1}\delta(\omega)\delta(\mathbf{q})&=\int\mathrm{d}t\mathrm{d}^{d}\mathbf{x}\,\mathrm{e}^{-\mi\omega t-\mi\mathbf{q}\cdot\mathbf{x}} \\
(2\pi)^{d+1}\delta(t)\delta(\mathbf{x})&=\int\mathrm{d}\omega\mathrm{d}^{d}\mathbf{q}\,\mathrm{e}^{\mi\omega t+\mi\mathbf{q}\cdot\mathbf{x}}.
\end{align}

A vectorial gaussian white noise with zero mean satisfies $\langle f_i(t,\mathbf{x})f_j(t',\mathbf{x}')\rangle=\Delta\,\delta_{ij}\delta(t-t')\delta(\mathbf{x}-\mathbf{x}')$ in real space and $\langle\tilde{f}_i(\omega,\mathbf{q})\tilde{f}_j(\omega',\mathbf{q}')\rangle=\Delta\,\delta_{ij}\delta(\omega+\omega')\delta(\mathbf{q}+\mathbf{q}')$ in Fourier space.

\section{Density-density and velocity-density gradient correlations}
\label{app:correlation}

In the following we compute the density-density and velocity-density gradient correlations for fluctuations around the homogenous state for the dynamic equations \eqref{eq:continuityEquation}-\eqref{eq:forcePersistent} with isotropic gaussian noise. Deviatoric quantities from the homogeneous state $n=n_0$ and $\mathbf{v}=\mathbf{0}$ are written without $\delta$ for brevity. In Fourier space, the linearized equations for these quantities are 
\begin{align}
(\mi\omega+k_d+D q^2)\tilde{n}&=-\mi n_0 q_i\tilde{v}_i \\
-(\eta q^2+\xi)\tilde{v}_i-(\bar\eta-\eta)q_i(q_j\tilde{v}_j)&=-\mi\bar{\alpha}q_i\tilde{n}+\tilde{f}_i \\
(\mi\omega\tau_p+1)\tilde{f}_i&=\tilde{\theta}_i
\end{align}
with $\bar{\eta}=\tilde{\eta}+2(d-1)\eta/d$. Combining the first two equations gives
\begin{widetext}
\begin{align}
\tilde{n}&=\frac{\mi n_0 q_i\tilde{f}_i}{(\mi\omega+k_d+Dq^2)(\bar{\eta}q^2+\xi)-\bar{\alpha}n_0 q^2}  \\
\tilde{v}_i
&=\frac{(q_iq_j-q^2\delta_{ij})\tilde{f}_j}{q^2(\eta q^2+\xi)}
-\frac{(\mi\omega+k_d+D q^2)(q_iq_j/q^2)\tilde{f}_j}{(\mi\omega+k_d+Dq^2)(\bar{\eta}q^2+\xi)-\bar{\alpha}n_0 q^2}.
\end{align}
\end{widetext}

We are interested in the density auto-correlation function $\langle\tilde{n}\tilde{n}'\rangle$ and the velocity-density gradient cross-correlation $\langle\tilde{v}_j\,\mi q_j'\tilde{n}'\rangle$, where the primes indicate quantities taken at wave-vector $\mathbf{q}'$ and time $t'$. We then obtain
\begin{widetext}
\begin{align}
\langle\tilde{n}\tilde{n}'\rangle &=\frac{\Delta n_0^2\,q^2}{|\mi\tau_p\omega+1|^2}\,\frac{\delta(\omega+\omega')\delta(\mathbf{q}+\mathbf{q}')}{|(\mi\omega+k_d+Dq^2)(\bar{\eta}q^2+\xi)-\bar{\alpha}n_0 q^2|^2}\\
\langle\tilde{v}_j\,\mi q_j'\tilde{n}'\rangle &=(\mi\omega+k_d+Dq^2)\langle\tilde{n}\tilde{n}'\rangle/n_0.
\end{align}
In real space this gives
\begin{align}
\langle\delta n^2\rangle&=\frac{\pi\,\Delta n_0^2}{(2\pi)^{d+1}}\int\mathrm{d}^d\mathbf{q}\
\frac{q^2}{-s(\mathbf{q})(\bar\eta q^2+\xi)^2[1-s(\mathbf{q})\tau_p]}, \\
\langle\mathbf{v}\cdot\bm\nabla n\rangle&=\frac{\pi\,\Delta n_0}{(2\pi)^{d+1}}\int\mathrm{d}^d\mathbf{q}\,
\frac{q^2(k_d+D q^2)}{-s(\mathbf{q})(\bar\eta q^2+\xi)^2[1-s(\mathbf{q})\tau_p]}.
\end{align}
\end{widetext}
Here, the growth rate $s(\mathbf{q})$ from~\eq{eq:growthExponent} has been introduced, which is negative for $\bar\alpha<\bar\alpha_c$. Both correlation functions are positive-definite in this case.

\section{Density-dependent force correlation}
\label{app:anisotropiccorrelation}

We show here that the nonlinear coupling between density and fluctuating force, in \eq{eq:forcePersistentAniso}, corresponds to a non-gaussian noise, and that a non-zero average polar force $\langle\mathbf f\rangle$ emerges if nonlinear advection of $\mathbf{f}$ is taken into account.

We first neglect the nonlinearity from advection in \eq{eq:forcePersistentAniso} to focus on the force-density coupling terms.

Introducing $\bm\Gamma=(\omega,\mathbf{q})$, the Fourier transform of \eq{eq:forcePersistentAniso} gives
\begin{align}\label{eq:forcePersistentAniso1F1}
(\mi\omega\tau_p+1)\tilde{f}_i(\bm\Gamma)&=\tilde{\theta}_i(\bm\Gamma)+\frac{1}{{(2\pi)^{(d+1)/2}}}\int\mathrm{d}\bm\Gamma'\,\tilde{F}_i(\bm\Gamma,\bm\Gamma') \\ \nonumber
\text{with}\quad\tilde{F}_i(\bm\Gamma,\bm\Gamma')&=(\epsilon_0\delta_{ij}-\epsilon_1 q_i'q_j')\tilde{n}(\bm\Gamma')\tilde{f}_j(\bm\Gamma-\bm\Gamma').
\end{align}
Using linearized equations for force balance and density dynamics, one has $\tilde{n}(\bm\Gamma)=N(\bm\Gamma)q_l\,\tilde{f}_l(\bm\Gamma)$, where $N(\Gamma)$ is a function of $\omega$ and $|\mathbf{q}|$. This implies that the integrand on the RHS of \eq{eq:forcePersistentAniso1F1} can be written as $\tilde{F}_i(\bm\Gamma,\bm\Gamma')=(\epsilon_0\delta_{ik}-\epsilon_1 q_i'q_k')N(\bm\Gamma')q'_l\tilde{f}_k(\bm\Gamma-\bm\Gamma')\tilde{f}_l(\bm\Gamma')$. Thus, the mean polar force reads
\begin{align}
\langle\tilde{f}_i\rangle=\int\mathrm{d}\bm\Gamma'\,\frac{N(\bm\Gamma')(\epsilon_0\delta_{ik}-\epsilon_1 q'_iq'_k)q'_l}{{(2\pi)^{(d+1)/2}}(\mi\omega\tau_p+1)}\langle\tilde{f}_k(\bm\Gamma-\bm\Gamma')\tilde{f}_l(\bm\Gamma')\rangle
\end{align}

To obtain the force correlations at lowest order, one can use an iterative approach by replacing $\tilde{f}_i(\bm\Gamma)\simeq\tilde{\theta}_i(\bm\Gamma)/(\mi\omega\tau_p+1)$ in the integral on the RHS of~\eq{eq:forcePersistentAniso1F1}, and assuming no correlation between force and density such that $\tilde{n}(\bm\Gamma')\simeq\langle\tilde{n}\rangle$ as in~\cite{Vafa2021}. Then we get
\begin{widetext}
\begin{align}
(\mi&\omega\tau_p+1)\tilde{f}_i(\bm\Gamma)\simeq\tilde{\theta}_i(\bm\Gamma) 
+\frac{\langle\tilde{n}\rangle}{{(2\pi)^{(d+1)/2}}}\int\mathrm{d}\bm\Gamma''\,(\epsilon_0\delta_{ij}-\epsilon_1 q_i''q_j'')\frac{\tilde{\theta}_j(\bm\Gamma-\bm\Gamma'')}{\mi(\omega-\omega'')\tau_p+1}.
\end{align}
Taking the product of these equations to obtain the force correlations $\langle\tilde{f}_k(\bm\Gamma-\bm\Gamma')\tilde{f}_l(\bm\Gamma')\rangle$, this leads after integration over $\bm\Gamma''$ to
\begin{align}
\langle\tilde{f}_k(\bm\Gamma-\bm\Gamma')\tilde{f}_l(\bm\Gamma')\rangle
&\simeq\Delta\,\delta_{kl}\delta(\bm\Gamma) 
+\frac{\Delta\langle\tilde{n}\rangle\,(\epsilon_0\delta_{kl}-\epsilon_1 q_kq_l)(2-\mi\omega\tau_p)}{{(2\pi)^{(d+1)/2}}\,|C(\omega-\omega',\omega')|^2},
\end{align}
\end{widetext}
where $C(\omega,\omega')=(\mi\omega\tau_p+1)(\mi\omega'\tau_p+1)$. This shows that the nonlinear coupling terms between force and density renormalize the force fluctuations, where in particular the $\epsilon_1$-term induces anisotropic force fluctuations. Alternatively, one could directly postulate density-dependent force fluctuations $\langle\theta_i(\mathbf{r},t)\theta_j(\mathbf{r}',t')\rangle=\Delta\{[1+\epsilon_0\langle\delta n(\mathbf{r},t)\rangle]\delta_{ij}+\epsilon_1\langle\partial_i\partial_jn(\mathbf{r},t)\rangle\}\delta(\mathbf{r}-\mathbf{r}')\delta(t-t')$ instead of a deterministic coupling between the fluctuating force and density.

Using the parity properties of integration over $\mathbf{q}'$, we finally obtain
\begin{widetext}
\begin{align}
(\mi\omega\tau_p+1)\langle\tilde{f}_i\rangle 
&
=\int\mathrm{d}\bm\Gamma'\,N(\bm\Gamma')\frac{\epsilon_0\delta_{ik}-\epsilon_1 q_i'q_k'}{{(2\pi)^{(d+1)/2}}}q'_l\langle\tilde{f}_k(\bm\Gamma-\bm\Gamma')\tilde{f}_l(\bm\Gamma')\rangle \\ 
&=\frac{\Delta\langle\tilde{n}\rangle\epsilon_1^2 q_kq_l}{{(2\pi)^{d+1}}}\int\mathrm{d}\bm\Gamma'\,\frac{N(\bm\Gamma')\,q_i'q_k'q_l'}{{(1-\mi\omega'\tau_p)C(\omega-\omega',\omega')}}\\
&=0.
\end{align}
\end{widetext}
Thus, in absence of advection, one does not obtain any effect from the force-density couplings on the mean polar force.

Incorporating the effect of advection adds another nonlinearity, and the mean polar force reads
\begin{align}
(\mi\omega\tau_p+1)\langle\tilde{f}_i(\bm\Gamma)\rangle&=\frac{1}{{(2\pi)^{(d+1)/2}}}\int\mathrm{d}\bm\Gamma'\,\langle\tilde{F}_i(\bm\Gamma,\bm\Gamma')\rangle
\end{align}
with $\tilde{F}_i=-\mi\tau_p \tilde{v}_m(\bm\Gamma')(q-q')_m\,\tilde{f}_i(\bm\Gamma-\bm\Gamma')$.

Again using linearized equations for force balance and density dynamics in first approximation, one has $\tilde{v}_m(\bm\Gamma')=[A(\bm\Gamma')q'_mq'_l+B(\bm\Gamma')(q'_mq'_l-{q'}^2\delta_{ml})]\tilde{f}_l(\bm\Gamma')$, which implies that $\tilde{F}_i(\bm\Gamma,\bm\Gamma')=-\mi\tau_p[A(\bm\Gamma')q'_mq'_l+B(\bm\Gamma')(q'_mq'_l-{q'}^2\delta_{ml})](q-q')_m\,\tilde{f}_i(\bm\Gamma-\bm\Gamma')\tilde{f}_l(\bm\Gamma')$, so that
\begin{widetext}
\begin{align}
C(\omega-\omega',\omega')\langle\tilde{F}_i(\bm\Gamma,\bm\Gamma')\rangle 
&=-\mi\tau_p\,[A(\bm\Gamma')(q_mq'_m-{q'}^2)q'_l+B(\bm\Gamma')(q'_mq'_l-{q'}^2\delta_{lm})q_m] 
\left[\Delta\,\delta(\bm\Gamma)\delta_{il}
+\frac{\Delta\langle\tilde{n}\rangle\,(\epsilon_0\delta_{il}-\epsilon_1 q_iq_l)(2-\mi\omega\tau_p)}{{(2\pi)^{(d+1)/2}}\,C^*(\omega-\omega',\omega')}
\right].
\end{align}
\end{widetext}
The $B$-term vanishes by integration over $\omega'$, and, using the parity properties of integration over $\mathbf{q}'$,  only a term $\propto q_m\int\mathrm{d}^d\mathbf{q}'\,A(\bm\Gamma')q'_mq'_l$ remains. Explicitly, $q^2A(\bm\Gamma)=-(\mi\omega+k_d+Dq^2)/[(\mi\omega-s(\mathbf{q}))(\bar\eta q^2+\xi)]$ with the growth rate $s(\mathbf{q})$ from~\eq{eq:growthExponent}. Then, integration over $\omega'$ removes the $\epsilon$-independent fluctuations and one gets a finite mean polar force
\begin{equation}
\langle\tilde{f}_i\rangle=\frac{\Delta}{\xi(\bar\eta/\xi)^{d/2}}(\epsilon_0-\epsilon_1 q^2)\mi q_i\langle\tilde{n}\rangle\,I
\end{equation}
with $I$ a positive dimensionless integral. This corresponds to $\langle\mathbf{f}\rangle\sim(\Delta/\xi)(\bar\eta/\xi)^{-d/2}(\epsilon_0\bm\nabla n+\epsilon_1\bm\nabla\nabla^2 n)$.

For a $1$d system in the parametric limit $\bar\alpha,\tau_p\rightarrow 0$, we obtain explicitly $I=1/8$, which translates exactly into $\langle f_{1\rm D}\rangle=\Delta/(8\sqrt{\bar\eta\xi})\langle\epsilon_0\nabla n+\epsilon_1\nabla^3 n\rangle$. In this particular case, the only approximation originates from the iterative method used to derived the force fluctuations in Fourier space.

\bibliography{refs}

\end{document}